\documentclass[journal,twoside,web]{ieeecolor}
\usepackage{generic}
\usepackage{cite}
\usepackage{amsmath,amssymb,amsfonts}
\usepackage{algorithmic}
\usepackage{graphicx}
\usepackage{textcomp}
\usepackage[version=4]{mhchem}

\def\BibTeX{{\rm B\kern-.05em{\sc i\kern-.025em b}\kern-.08em
    T\kern-.1667em\lower.7ex\hbox{E}\kern-.125emX}}
    
\DeclareUnicodeCharacter{2212}{\ensuremath{-}}

\begin{document}

\title{Thickness Dependence of Coercive Field in Ferroelectric Doped-Hafnium Oxide}
\author{Revanth Koduru, \IEEEmembership{Student Member, IEEE}, and Sumeet K. Gupta, Jr., \IEEEmembership{Senior Member, IEEE}
\thanks{This work was supported by National Science Foundation (NSF).}
\thanks{We thank Dr. Martin M. Frank (IBM Research, Yorktown Heights, NY, 10598 USA) for his valuable discussions and insights. We also acknowledge Tanmoy Kumar Paul (Elmore Family School of Electrical and Computer Engineering, Purdue University, West Lafayette, IN 47906 USA) for providing the orthorhombic \ce{HfO2} crystal structure shown in Fig.~\ref{figure1}b.}
\thanks{Revanth Koduru is with the Elmore Family School of Electrical and Computer Engineering, Purdue University, West Lafayette, IN 47906 USA (email: kodurur@purdue.edu).}
\thanks{Sumeet K. Gupta is with the Elmore Family School of Electrical and Computer Engineering, Purdue University, West Lafayette, IN 47906 USA (email: guptask@purdue.edu).}}
\maketitle

\begin{abstract}
Ferroelectric hafnium oxide (${HfO_2}$) exhibits a thickness-dependent coercive field $(E_c)$ behavior that deviates from the trends observed in perovskites and the predictions of Janovec-Kay-Dunn (JKD) theory. Experiments reveal that, in thinner $HfO_2$ films ($<100\,nm$), $E_c$ increases with decreasing thickness but at a slower rate than predicted by the JKD theory. In thicker films, $E_c$ saturates and is independent of thickness. Prior studies attributed the thick film saturation to the thickness-independent grain size, which limits the domain growth. However, the reduced dependence in thinner films is poorly understood. In this work, we expound the reduced thickness dependence of $E_c$, attributing it to the anisotropic crystal structure of the polar orthorhombic (o) phase of $HfO_2$. This phase consists of continuous polar layers (CPL) along one in-plane direction and alternating polar and spacer layers (APSL) along the orthogonal direction. The spacer layers decouple adjacent polar layers along APSL, increasing the energy barrier for domain growth compared to CPL direction. As a result, the growth of nucleated domains is confined to a single polar plane in $HfO_2$, forming half-prolate elliptical cylindrical geometry rather than half-prolate spheroid geometry observed in perovskites. By modeling the nucleation and growth energetics of these confined domains, we derive a modified scaling law of $E_c \propto d^{-1/2}$ for $HfO_2$ that deviates from the classical JKD dependence of $E_c \propto d^{-2/3}$. The proposed scaling agrees well with the experimental trends in coercive field across various ferroelectric $HfO_2$ samples.
\end{abstract}


\section{Introduction}
\label{sec:introduction}

\IEEEPARstart{T}{he}  discovery of ferroelectricity in CMOS-compatible hafnium oxide (\ce{HfO2}) has gained significant attention \cite{boscke_ferroelectricity_2011}. Certain fabrication conditions or suitable dopants (such as \ce{Zr}, \ce{Y}, \ce{Si}, etc.) stabilize a non-centrosymmetric orthorhombic phase (o-phase) with \ce{{Pca2}_1} space group in \ce{HfO2}, which exhibits spontaneous polarization and FE behavior \cite{boscke_ferroelectricity_2011, schroeder_fundamentals_2022, muller_ferroelectricity_2012, materlik_origin_2015}. 
The o-phase crystal structure of \ce{HfO2} gives rise to FE properties distinct from conventional ferroelectrics such as perovskites. Notably, \ce{HfO2} exhibits scale-free ferroelectricity \cite{lee_scale-free_2020}, i.e., retains FE behavior down to near unit-cell thickness. This property coupled with CMOS process compatibility \cite{muller_ferroelectric_2013}, makes FE \ce{HfO2} a strong candidate for logic and memory applications in scaled technologies \cite{mikolajick_hafnium_2018,mikolajick_past_2020}. Given this, it is crucial to understand the impact of thickness scaling on the FE properties of \ce{HfO2}. This work focuses on the behavior of the coercive field $(E_c)$ as a function of FE \ce{HfO2} thickness $(d)$.

\begin{figure*}[!h]
\centerline{\includegraphics[width=\linewidth]{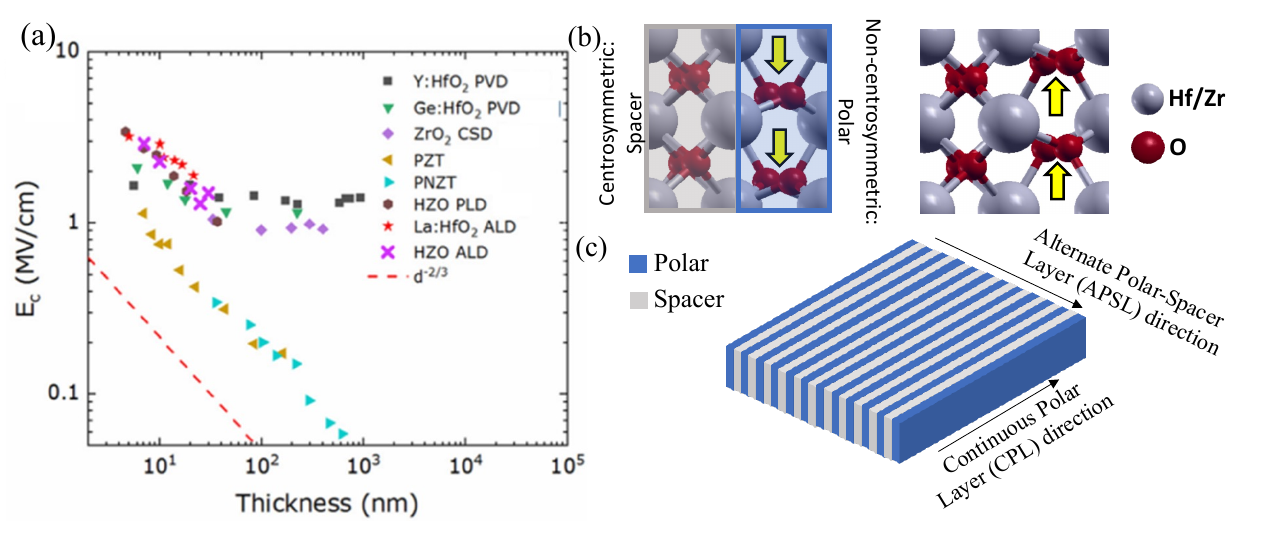}}
\caption{a) Experimental trends (reproduced from \cite{park_perspective_2023} with permission) of coercive field $(E_c)$ versus ferroelectric thickness $(d)$ for various samples of \ce{HfO2} and perovskites, showing the different regimes and the reduced thickness dependence in \ce{HfO2}. b) Unit lattice crystal structure of the orthorhombic phase of \ce{HfO2} in upward and downward polarization, highlighting the centrosymmetric (spacer) and non-centrosymmetric(polar) sublattices. c) 3D arrangement of o-\ce{HfO2} showing the anisotropic crystal structure with alternating polar-spacer layers (APSL) in one lateral direction and continuous polar layers (CPL) in the other.}
\label{figure1}
\end{figure*}

Experiments \cite{park_perspective_2023, materano_polarization_2020, mimura_thickness-dependent_2018, park_study_2015, toriumi_material_2019, migita_polarization_2018} show that thickness-dependence of $E_c$ in \ce{HfO2} differs from the perovskites \cite{park_perspective_2023,bjormander_thickness_1995,pertsev_coercive_2003}. In perovskites, $E_c$ follows a scaling behavior of $E_c\propto d^{-2/3}$, which is explained by the classical Janovec-Kay-Dunn (JKD) theory \cite{janovec_theory_1958, kay_thickness_1962, chandra_scaling_2004}. In contrast, the thickness dependence of $E_c$ in \ce{HfO2} can be divided into two regimes (Fig. ~\ref{figure1}a). In thinner films (typically below $100\,nm$), $E_c$ increases with decreasing $d$, but at a slower rate than perovskites. In thicker films, however, $E_c$ saturates and becomes independent of $d$.

Prior works \cite{park_perspective_2023, materano_polarization_2020, mimura_thickness-dependent_2018} have focused on explaining the saturation behavior in thicker films. While the FE thickness limits the polarization switching in perovskites \cite{janovec_theory_1958,kay_thickness_1962,chandra_scaling_2004, materano_polarization_2020}, it is proposed that in \ce{HfO2}, the limiting factor is the grain size or the intrinsic domain size \cite{mimura_thickness-dependent_2018,migita_polarization_2018,materano_polarization_2020}. Since the grain and domain sizes do not scale proportionally with film thickness in thicker films \cite{materano_polarization_2020}, they remain smaller than $d$. As a result, polarization switching is constrained by the grain or domain sizes in thicker films and $E_c$ remains largely constant with $d$. 

However, the physical origin of the weaker dependence of $E_c$ in thinner films remain poorly understood. In particular, $E_c$ increases slower than predicted by the JKD theory, even when the grain size is significantly larger than film thickness \cite{park_perspective_2023,park_surface_2017}. Understanding this behavior is crucial for optimizing thin \ce{HfO2} films in next-generation devices. 

In this work, we address this need by revisiting the JKD theory in the context of \ce{HfO2}. The JKD theory \cite{janovec_theory_1958, kay_thickness_1962} models  polarization switching as a two step process: nucleation of reverse domains followed by their growth. It assumes an isotropic in-plane growth, modeling domains as half-prolate spheroids. The film thickness constrains domain growth and thereby, influences the energy of the nucleus capable of switching the FE film, leading to the $E_c\propto d^{-2/3}$ dependence.

We propose that the isotropic domain assumption does not hold for \ce{HfO2} due to the anisotropic crystal structure of its orthorhombic (o) phase \cite{lee_scale-free_2020,paul_direction-dependent_2022,materlik_origin_2015,park_atomic-scale_2024} (Fig.~\ref{figure1}b). The polar o-phase exhibits structural anisotropy in the lateral (in-plane) directions, featuring continuous polar layers (CPL) along one in-plane direction and alternating polar-spacer layers (APSL) along the orthogonal direction (Fig.~\ref{figure1}c). The spacer layers increase the energy barrier for domain growth along APSL compared to CPL direction. This confines the domain growth and in turn, the nucleated domains to a single polar layer.

Due to the anisotropic nature of \ce{HfO2}, the domain geometry deviates from the half-prolate spheroid considered in the JKD theory. Instead, we model the domains in \ce{HfO2} as half-prolate elliptical cylinders (Fig.~\ref{figure2}a), reflecting the confined in-plane growth. Following an approach similar to the JKD theory, we analyze the nucleation and growth energetics for this geometry and derive a modified thickness dependence of $E_c$ in \ce{HfO2}:
\begin{equation}
    E_c \propto d^{-1/2}
\end{equation}
This reduced exponent of 1/2, compared to 2/3 of the JKD theory, matches well with experimental data of FE \ce{HfO2} films, as discussed in Section \ref{sec:exp_comp}.

It is important to note that some experimental studies, particularly on epitaxial \ce{HfO2} films \cite{fina_epitaxial_2021, lyu_growth_2019, song_epitaxial_2020}, report close to the JKD scaling. We believe this is due to the dominance of a different polar phase, namely the rhombohedral (R3mn) phase \cite{wei_rhombohedral_2018, petraru_distinguishing_2024}.  R3mn phase lacks the APSL structure of the o-phase and hence, does not exhibit anisotropic behavior. This work focuses on the more common \ce{HfO2} films (e.g. fabricated by processes such as ALD) with the dominant o-phase.

The key contributions of this work include:
\begin{itemize}
    \item A physical understanding of the experimentally observed weaker dependence (compared to that predicted by the JKD theory) of coercive field $(E_c)$ on thickness $(d)$ in FE \ce{HfO2} at low film thicknesses.
    \item Analysis of the energetics of domain nucleation in o-\ce{HfO2}, incorporating the confined growth and the resultant half-prolate elliptical cylindrical domain geometry.
    \item Derivation of a modified scaling relation for the coercive field, showing $E_c \propto d^{-1/2}$, which aligns with the experimental observations in \ce{HfO2} (but deviates from the standard $d^{-2/3}$ scaling of the JKD theory).
\end{itemize}

\section{Domain Structure in Ferroelectric \ce{HfO2}}
\label{sec:domain_structure}
Polarization switching in ferroelectrics is widely understood to initiate with the nucleation of a reverse domain, followed by its growth \cite{tagantsev_non-kolmogorov-avrami_2002, park_perspective_2023, materano_polarization_2020}. Perovskites exhibit isotropic in-plane domain growth, forming half-prolate spheroid domains --- an assumption central to the JKD theory \cite{janovec_theory_1958, kay_thickness_1962}. In contrast, o-phase of \ce{HfO2} exhibits anisotropic crystal structure, leading to direction-dependent domain growth in the in-plane direction. As a result, the assumptions of the domain geometry and the derivation of the thickness-dependence of coercive field must be revisited for \ce{HfO2}.

The unit cell of o-\ce{HfO2} consists of \ce{Hf}-atoms at the corners and face centers of the lattice, with four \ce{O}-atoms occupying positions within the crystal lattice. This structure can be viewed as comprising two sublattices: centrosymmetric and non-centrosymmetric (Fig.~\ref{figure1}b) \cite{lee_scale-free_2020, paul_direction-dependent_2022}. In the centrosymmetric sublattice, \ce{O}-atoms are symmetric relative to the surrounding \ce{Hf} atoms, resulting in non-polar or dielectric behavior. In the non-centrosymmetric sublattice, \ce{O}-atoms are displaced from their symmetric positions, giving rise to the spontaneous polarization and FE properties.

In the 3D \ce{HfO2} crystal lattice, polar sublattices align with other polar sublattices along the out-of-plane (vertical) direction. In the same way, spacer sublattice align with other spacer sublattices. Hence, distinct polar and spacer layers are formed (Fig.~\ref{figure1}c). However, the in-plane arrangement is anisotropic. Along one in-plane direction, referred to as alternating polar spacer layer (APSL) direction, polar and spacer layers alternate. Along the orthogonal in-plane direction, referred to as continuous polar layer (CPL) direction, either polar or spacer layers extend continuously \cite{lee_scale-free_2020, paul_direction-dependent_2022, park_atomic-scale_2024}.

First principles studies \cite{lee_scale-free_2020, paul_direction-dependent_2022} have shown that, along the APSL direction, spacer layers act as decoupling planes between adjacent polar layers. This leads to direction-dependent energy barriers for domain growth with a higher barrier along the APSL direction compared to the CPL direction \cite{paul_direction-dependent_2022}. As a result, domain propagation is energetically favored along the CPL while strongly suppressed along the APSL direction. This anisotropic barriers confines the nucleated domains to within a single polar layer along APSL, a behavior supported by first-principles calculations \cite{lee_scale-free_2020, paul_direction-dependent_2022} and experimental observations \cite{park_atomic-scale_2024} of unit-cell-wide domains.

\begin{figure}[!t]
\centerline{\includegraphics[width=\columnwidth]{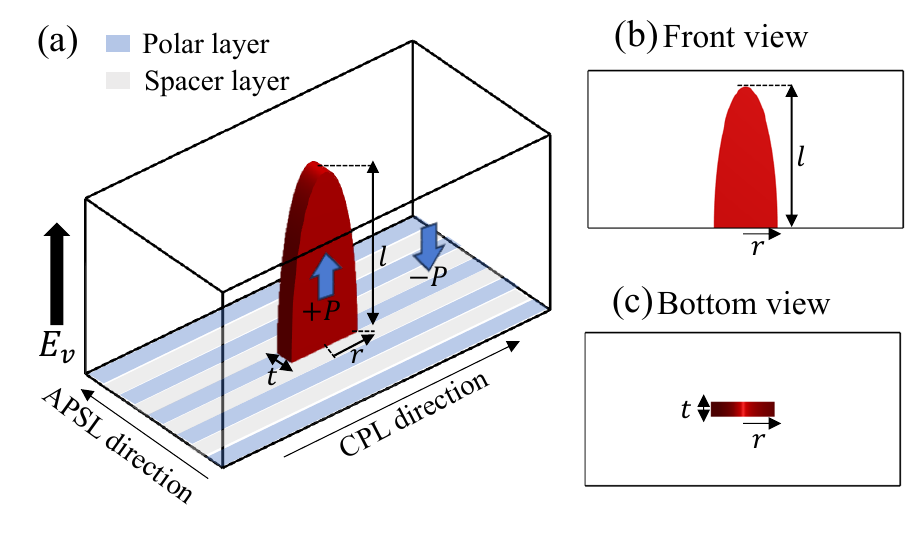}}
\caption{Structure of the half-prolate elliptical cylinder reverse domain with upward polarization $(+P)$ confined to a single polar layer along the APSL direction in a fully poled $-P$ polarized \ce{HfO2} layer in a) 3D view, b) 2D front-view and c) 2D bottom view.}
\label{figure2}
\end{figure}

To account for this confinement of domains along polar layer in \ce{HfO2}, we model the nucleated domains as half-prolate elliptical cylinder with dimensions $(r,l,t)$, as shown in Fig.~\ref{figure2}a. The elliptical face extends vertically into the FE \ce{HfO2} film along the major axis $(l)$ while the minor axis $(r)$ lies along the FE-metal/dead-layer/dielectric interface. It is noteworthy that $l\gg r$ (Fig.~\ref{figure2}b). Since the domain is confined to a single polar layer along the APSL (as shown in Fig.~\ref{figure2}a,c), the thickness $(t)$ of the elliptical cylinder is taken to be half the lattice constant of \ce{HfO2}, consistent with the domain widths observed in first-principles simulations \cite{paul_direction-dependent_2022} along the APSL direction. For simplicity, we assume that the elliptical cross section remains constant along the domain thickness.

\section{Energy Landscape of Reverse Domains in \ce{HfO2}}

For polarization switching via nucleation and growth of a reversal domain, the process must be energetically favorable at an applied external electric field. To evaluate the conditions necessary for such switching, we consider a uniform o-\ce{HfO2} FE layer of thickness $(d)$ in a metal-ferroelectric-metal (MFM) capacitor structure. To focus on the intrinsic switching behavior, we neglect the effects of polycrystallinity, polymorphism and defects. 

In a real FE sample, nucleation can occur from various sites such as grain boundaries, structural imperfections, or metal-ferroelectric interfaces. For simplicity, we assume that nucleation initiates from the metal-ferroelectric interface. Given that the reverse domain is confined to a single polar layer along APSL, each nucleation and subsequent growth can switch only a single polar layer. However, switching of the entire FE film involves multiple such nucleation events across different polar layers. We assume that these nucleation events are independent of each other and focus on one such nucleation event and subsequent growth.

Additionally, the non-uniformity of the polycrystalline FE films lead to spatial variations in local electric fields and nucleation thresholds \cite{koduru_phase-field_2023}. However, we restrict our analysis to a single-grain scenario, where the polarization axis is aligned with the film thickness and coincides with the direction of the electric field. This single grain approach can be extended to polycrystalline films by accounting for the fact that only a component of electric field aligned with the local polarization axis of each grain contributes to the switching.

Without any loss of generality, we consider the FE layer to be initially in a fully poled state of downward polarization $(-P_S)$. The metal electrodes are biased to create an upward electric field $(E_v)$ (Fig.~\ref{figure2}a). We consider this initial configuration to be our reference energy state $U_{ref}$. Now, consider the nucleation of an reverse $(+P_s)$ domain, modeled as half-prolate elliptical cylinder  (Fig.~\ref{figure2}a), resulting in new energy state $U_{nuc}$. For nucleation to be energetically favorable, the associated change in energy $(U_\Delta)$ must be negative:
\begin{equation} \label{nucl_condition}
    U_\Delta = U_{nuc} - U_{ref} < 0
\end{equation}
Further, for this domain to switch the polar layer, its growth must also be energetically favorable. That is, any increase $(dV>0)$ in domain volume $(V)$ must lead to a decrease in energy $(dU_\Delta<0)$. Therefore, we define the coercive field $(E_c)$  as the minimum external electric field required for 
\begin{enumerate}
    \item the nucleation of a reverse domain to be energetically favorable, and
    \item its subsequent growth to be energetically favorable, 
\end{enumerate}
leading to polarization reversal of the FE layer.

To derive $E_c$, we analyze different energy components of $U_\Delta$ associated with the half-prolate elliptical cylinder reverse $(+P_s)$ domain with dimensions $(r,l,t)$, where $l\gg r$. This domain experiences a total electric field which is the vector sum of two components:
\begin{itemize}
    \item Homogeneous electric field due to the applied voltage across the electrodes, assuming a uniform polarization in the FE layer $(\nabla.P=0)$. This is referred to as the externally applied electric field $(E_v)$, consistent with the JKD theory \cite{janovec_theory_1958}.
    \item Inhomogeneous electric field $(E_i)$ arising from spatial variations in polarization $(\nabla.P\neq 0)$, particularly near domain walls. This is similar to depolarization field in the JKD theory \cite{janovec_theory_1958}. However, we call this inhomogeneous field to avoid confusion with the depolarization field arising from finite screening of metal electrodes considered in Section \ref{sec:exp_comp}.
\end{itemize}

The change in the electrostatic energy $(U_{\Delta e})$ associated with the $+P$ domain due to the influence of $E_v$ is given by:
\begin{equation} \label{elec_static_energy}
    U_{\Delta e} = -2E_vP_sV
\end{equation}
where $V$ is the volume of the half-prolate elliptical cylinder domain given by:
\begin{equation}
    V = \frac{\pi rlt}{2}
\end{equation}

The inhomogeneous electric field $(E_i)$ causes a change in the electrostatic displacement $(D)$ by $D_{\Delta i}$. The associated change in energy is given by
\begin{equation}
    U_{\Delta i} = \int_V E_iD_{\Delta i}dV
\end{equation}
Evaluating this term requires one to derive the expressions for inhomogeneous electric field $E_i$ and the resultant $D_{\Delta i}$. Instead of deriving the expressions from scratch, we utilize the inhomogeneous energy formulation derived by Landauer \cite{landauer_electrostatic_1957} for 2D half-elliptical domains and extend it to 3D half-prolate elliptical cylinder geometry, giving us:
\begin{equation} 
    U_{\Delta i} = \frac{P_s^2r^2lt}{4\varepsilon_c\left(\sqrt{\frac{\varepsilon_a}{\varepsilon_c}}l+r\right)}
\end{equation}
where, $\varepsilon_a$ and $\varepsilon_c$ are the in-plane and out-of-plane permittivity values of the FE respectively. For the case of $l\gg r$, this simplifies to
\begin{equation} \label{inhom_energy}
    U_{\Delta i} = \frac{P_s^2r^2t}{4\sqrt{\varepsilon_a\varepsilon_c}}
\end{equation}

Further, the nucleated domain also incurs a surface energy cost $(U_{\Delta s})$ due to the domain wall formation:
\begin{equation} \label{surf_energy}
    U_{\Delta S} = \sigma_eS_e + \sigma_cS_c
\end{equation}
here, $\sigma_c$ is the energy density of the 180$^\circ$ domain wall and $S_c$ is the lateral surface area of the elliptical cylinder. Using Ramanujan's approximation for the perimeter of an ellipse, we get $S_c$ as:
\begin{equation} \label{Sc}
    S_c = \frac{\pi}{2}(r+l)t\left(1+\frac{3h}{10+\sqrt{(4-3h)}}\right)
\end{equation}
where $h=(l-r)^2/(l+r)^2$. For $l\gg r$, we have $h\approx1$ and \eqref{Sc} simplifies to:
\begin{equation}
    S_c = \frac{7\pi}{11}lt
\end{equation}
The term $\sigma_e$ denotes the energy density of the domain wall surface with the spacer layer whose surface area is given by:
\begin{equation}
    S_e = 2\pi rl
\end{equation}

Additionally, the domain must overcome a certain energy barrier ($U_h$) associated with the polarization reversal. This barrier can be expressed in terms of a critical electric field ($E_h$), analogous to the formulation in the JKD theory, as:
\begin{equation} \label{barrier_energy}
    U_h = 2E_hP_sV = E_hP_s\pi rlt
\end{equation}
This energy barrier must be supplied by the external electric field and is therefore included as part of $U_\Delta$. 

Combining these different energy terms (\eqref{elec_static_energy} to \eqref{barrier_energy}), we get $U_\Delta$ associated with the $+P_s$ domain as:
\begin{multline} \label{energy_expanded}
    U_\Delta(E_v,r,l) = -E_vP_s\pi trl + \frac{7\pi\sigma_ct}{11}l + 2\sigma_e\pi rl \\
    + \frac{P_s^2r^2t}{4\sqrt{\varepsilon_a\varepsilon_c}}r^2 + E_hP_s\pi trl
\end{multline} 

This is a function of the external electric field ($E_v$) and the nucleus dimensions $(r,l)$ with domain thickness $t$ treated as a constant (as discussed in Section~\ref{sec:domain_structure}). Combining the constants \eqref{energy_expanded} can be further simplified to:
\begin{equation} \label{U_eq}
    U_\Delta(E_{va},r,l) = -aE_va.r.l + b.l+c.r^2
\end{equation}
where
\begin{align}
    a &= P_s\pi t\\
    b &= \frac{7\pi\sigma_ct}{11} \\
    c &= \frac{Ps^2t}{4\sqrt{\varepsilon_a\varepsilon_c}}\\
\text{and}\\
    E_{va} &= E_v-E_h-\frac{2\sigma_e}{P_s} \label{E_va}
\end{align}
$E_{va}$ is the apparent electric field and rest of the analysis is carried out in terms of $E_{va}$

Since the definition of $E_c$ requires both nucleation and domain growth to be energetically favorable, we will break our analysis into two steps. First, in Section~\ref{sec:growth_energy}, we establish the conditions under which the growth of an already nucleated domain is energetically favorable. Next, in Section~\ref{sec:nucleation_energy}, we identify the conditions under which such a domain can nucleate. Combining together, we derive the scaling relation between $E_c$ and $d$.

\section{Conditions for Growth of Reverse Domains in \ce{HfO2}}
\label{sec:growth_energy}

\begin{figure*}[!t]
\centerline{\includegraphics[width=\linewidth]{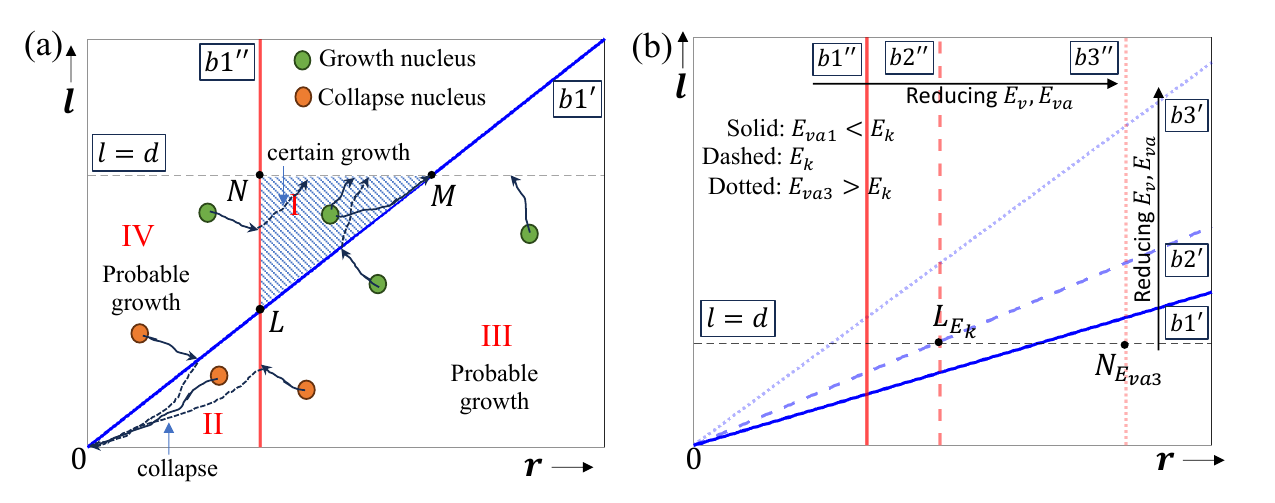}}
\caption{a) Plot of boundary curves \eqref{r_growth_b1} and \eqref{l_growth_b1}, which divide the $l-r$ plane into 4 regions based on energetic favorability of domain growth. Representative trajectories of nucleated domains in each region illustrated to show possible growth or collapse behaviors. (b) Evolution of boundary curves with decreasing external electric field $E_v$ (and consequently  $E_{va}$), highlighting critical field $E_{k}$ below which region I, representing certain growth, disappears.}
\label{figure3}
\end{figure*}

Assuming the reverse domain has already nucleated, its growth is energetically favorable only if an increase in its volume by $dV$ decreases the energy i.e. $dU_\Delta/dV<0$. Among the different energy components, the electrostatic energy $(U_{\Delta e})$ favors the domain growth as $U_e$ decreases with increasing $V$ (see \eqref{elec_static_energy}). This occurs because the polarization of the nucleated domain is aligned with the applied electric field. In contrast, the other energy components --- energy due to inhomogeneous electric field $(U_{\Delta i})$, surface $(U_{\Delta s})$ and barrier $(U_{\Delta h})$ energies --- oppose the formation and growth of the domain (\eqref{inhom_energy}, \eqref{surf_energy}, \eqref{barrier_energy}). To determine, whether growth is favorable under these competing effects, we evaluate the condition $dU_\Delta/dV<0$ in terms of nucleus dimensions $r,l$ at a given $E_{va}$.

For a nucleus to grow in $r$, the condition $\partial U_\Delta/\partial r < 0$ must be satisfied which gives:
\begin{equation} \label{r_growth}
    \frac{\partial U_\Delta}{\partial r} = -aE_{va}l + 2cr<0 \\
    \implies l > \frac{2cr}{aE_{va}}
\end{equation}
The associated boundary curve is given by
\begin{equation} \label{r_growth_b1}
    l=\frac{2cr}{aE_{va}}
    \tag{b1'}
\end{equation}

For nucleus to grow in $l$, the condition to be satisfied is $\partial U_\Delta/\partial l<0$, which gives
\begin{equation} \label{l_growth}
    \frac{\partial U_\Delta}{\partial l} = -aE_{va}r + b <0\\
    \implies r > \frac{b}{aE_{va}}
\end{equation}
The corresponding boundary curve is 
\begin{equation} \label{l_growth_b1}
    r = \frac{b}{aE_{va}} \tag{b1''}
\end{equation}

The boundary curves (\eqref{r_growth_b1} and \eqref{l_growth_b1}) define the critical thresholds for domain growth with respect to nucleus dimensions $r,l$. We plot these curves on $l-r$ plane (Fig. \ref{figure3}a) with $l$ constrained by the physical FE thickness $d$. Three points of intersection on this $l-r$ plot are of interest:
\begin{itemize}
    \item Point $L$ (referred to as $L_{E_{va}}$ to stress its dependence on $E_{va}$ ): This is the intersection of \eqref{r_growth_b1} and \eqref{l_growth_b1} and its coordinates are:
    \begin{equation} \label{L_coord}
        (r_{L,E_{va}},l_{L,E_{va}}) = \left(\frac{b}{aE_{va}},\frac{2cb}{a^2}\frac{1}{E_{va}^2}\right)
    \end{equation}
    
    \item Point $M$ (or $M_{E_{va}}$): the intersection of \eqref{r_growth_b1} and horizontal line $l=d$. The coordinates are given by:
    \begin{equation} \label{M_coord}
    (r_{M,E_{va}},l_{M,E_{va}}) = \left(\frac{adE_{va}}{2c},d\right)
    \end{equation}
    
    \item Point $N$ (or $N_{E_{va}}$): the intersection of \eqref{l_growth_b1} and horizontal line $l=d$ with coordinates:
    \begin{equation} \label{N_coord}
        (r_{N,E_{va}},l_{N,E_{va}}) = \left(\frac{b}{aE_{va}},d\right)
    \end{equation}
\end{itemize}

This $l-r$ plot can be divided into four regions based on the favorability of the domain growth (Fig.~\ref{figure3}a):
\begin{itemize}
    \item Region-I:
    \begin{equation}
        \partial U_\Delta/\partial r<0, \partial U_\Delta/\partial l< 0
    \end{equation} 
    In this region, nuclei grow both in $r$ and $l$, indicating energetically favorable domain growth. Although, the precise trajectory of growth depends on the relative growth rate of $r$ and $l$ (which is beyond the scope of this work), the domain will ultimately reach the boundary $l=d$ (Fig.~\ref{figure3}a) i.e. span the full FE thickness. 
    
    Once the reverse domain reaches the opposite metal electrode, polarization charges at the interface starts getting compensated by the metal, reducing $E_i$. The domain starts changing shape due to the growth only in $r$ with no growth in $l$ and hence, the equations derived here (which assume elliptical cylindrical shape) are no longer valid. However, due to the continued reduction of opposing force $E_i$, the domain continues to expand laterally in $r$, switching the polar layer \cite{kay_thickness_1962}. It is important to mention that our goal here is not to capture how the domain grows, but only to derive the favorable conditions for domain growth (after nucleation), which enables us to define $E_c$. 
    \item Region-II:
    \begin{equation}
        \partial U_\Delta/\partial r>0, \partial U_\Delta/\partial l > 0
    \end{equation}
    In this region, the growth of nucleus is energetically unfavorable. Nuclei shrink in both $r$ and $l$, ultimately collapsing (Fig.~\ref{figure3}a).
    \item Region-III:
    \begin{equation}
        \partial U_\Delta/\partial r>0, \partial U_\Delta/\partial l < 0
    \end{equation}
    Nuclei in this region shrink in $r$ but grow in $l$. This is a region of probable growth as the result depends on the trajectory of the nucleus growth (Fig.~\ref{figure3}a):
    \begin{itemize}
        \item If it first crosses the blue boundary \eqref{r_growth_b1}, it enters region-I and grows in $r$ and $l$.
        \item If it reaches the red boundary \eqref{l_growth_b1}, it enters region-II and collapses.
        \item If it reaches $l=d$ boundary, $E_i$ starts decreasing. At that point, the growth in $r$ becomes favorable again \cite{kay_thickness_1962}, enabling lateral expansion and eventual switching.
    \end{itemize}
    \item Region-IV:
    \begin{equation}
        \partial U_\Delta/\partial r<0, \partial U_\Delta/\partial l > 0
    \end{equation}
    Here, nuclei shrink in $l$ but grow in $r$. As region-III, this is a probable growth region with the outcome depending on the growth trajectory:
    \begin{itemize}
        \item If it intersects the red boundary \eqref{l_growth_b1}, nucleus enters region-I and grows in $r$, $l$.
        \item If it cross the blue boundary \eqref{r_growth_b1}, nucleus enters region-II and collapses.
    \end{itemize}
\end{itemize}

For sufficiently large $E_{va}$, region-I corresponds to guaranteed (certain) domain growth and regions III and IV represent conditions of probable growth. Since our objective is to determine $E_c$ (which can be loosely considered as threshold separating switching and non-switching regions), we analyze how reducing $E_v$, and consequently $E_{va}$ affects the regions in the $l-r$ plot. 

As $E_{va}$ reduces, slope of \eqref{r_growth_b1} increases and \eqref{l_growth_b1} shifts rightward (Fig.~\ref{figure3}b). There exists a critical field $E_{va} = E_k$, where the curves transform to (b2') and (b2''), whose intersection $L_{E_k}$ lies on the $l=d$ line (Fig.~\ref{figure3}b). At this point, region-I, the region of certain domain growth vanishes. When $E_{va}$ reduces below $E_k$ to a value say $E_{va3}$, the boundary curves transform to (b3') and (b3'') (Fig.~\ref{figure3}b). These curves do not intersect in the physically permissible space (below $l=d$). In this scenario, only regions III and IV associated with probable growth remain. In addition, one can observe that as $d$ increases, the critical field $E_k$ required to sustain the region of certain growth reduces, indicating a stronger favorability of domain growth in thicker films at lower external field.

\section{Conditions for Nucleation and Thickness Dependence of Coercive Field}
\label{sec:nucleation_energy}

\begin{figure}[!t]
\centerline{\includegraphics[width=\columnwidth]{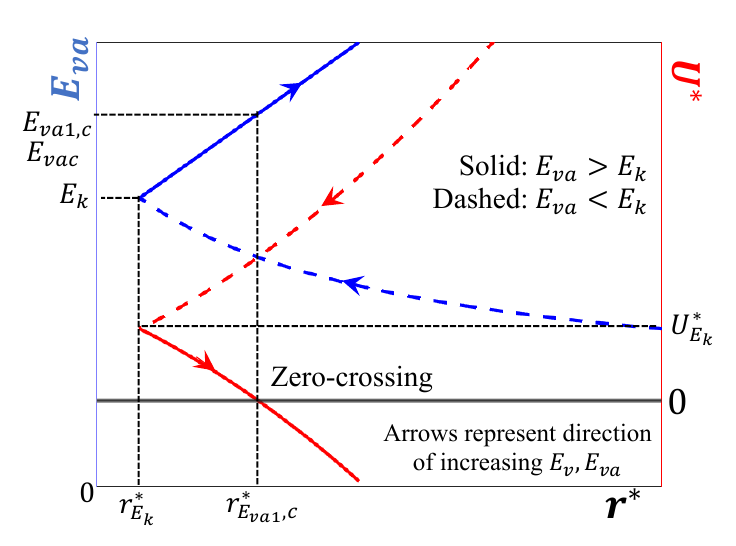}}
\caption{Apparent electric field $(E_{va})$ and the corresponding minimum energy of the nucleus $(U^*_{E_{va}})$ plotted against the radius $(r^*_{E_{va}})$ of the minimum-energy nucleus. Highlights the critical field $E_{vac}$, defined by the zero-crossing of $U^*_{E_{va}}$, marking the onset of energetically favorable nucleation.}
\label{figure4}
\end{figure}

In the previous section, we examined the size constraints that a nucleus must satisfy for domain growth to be energetically favorable at a given $E_v$ or $E_{va}$. Let us now turn our attention to answer the question whether the formation of such a nucleus is itself energetically favorable. We will do so by analyzing the energetics of nucleating a reverse domain that is also capable of growing. 

From \eqref{nucl_condition}, nucleation is energetically favorable if $U_\Delta$ is negative for a given $E_{va},r,l$. Since our goal is to determine $E_c$, we focus on finding the critical apparent field $(E_{va,c})$ at which nucleation first becomes favorable. Specifically, we will evaluate the minimum energy $U^*_{\Delta,E_{va}}$ across all nuclei sizes capable of growth as a function of $E_{va}$. Let the dimensions of the nucleus at this energy minimum be $(r^*_{E_{va}},l^*_{E_{va}})$. The value of $E_{va}$ needed at which $U^*_{\Delta,E_{va}}$ first becomes negative defines $E_{va,c}$. We can get $E_c$ from $E_{va,c}$ using \eqref{E_va} as follows
\begin{equation} \label{E_c_E_vac}
    E_c = E_{va,c} + E_h + \frac{2\sigma_e}{P_s}
\end{equation}

Let us consider different scenarios for $E_{va}$ to figure out the favorable conditions for nucleation:
\begin{enumerate}
    \item  We begin by considering the case where $E_{va}=E_k$. Here the minimum energy point corresponds to $L_{E_k}$ (Fig.~\ref{figure3}b), whose coordinates are given by substituting $E_{va}=E_k$ in \eqref{L_coord}:
    \begin{equation} \label{L_Ek_coord}
        (r^*_{E_k},l^*_{E_k}) = \left(\frac{b}{aE_k}, \frac{2cb}{a^2E_k^2}\right)
    \end{equation}
    Since $l^*_{E_k}=d$, we get (from $l^*_{E_k}$ in \eqref{L_Ek_coord}) the following relation for $E_k$:
    \begin{equation} \label{E_k}
        d=\frac{2cb}{a^2E_k^2} \implies E_k = \sqrt{\frac{2cb}{a^2d}}
    \end{equation}
    And, substituting \eqref{E_k} in $r^*_{E_k}$ from \eqref{L_Ek_coord}, we get:
    \begin{equation} \label{r_E_k}
        r^*_{E_k} = \sqrt{\frac{bd}{2c}}
    \end{equation}
    The energy $U^*_{\Delta E_k}$ at $(r^*_{E_k},l^*_{E_k},E_k)$ is obtained by substituting \eqref{E_k},\eqref{r_E_k}, and $l^*_{E_k}=d$ in \label{U_eq}:
    \begin{equation} \label{U_Ek}
        U^*_{\Delta,E_k} = \frac{bd}{2}
    \end{equation}
    Since this value is positive (i.e. the energy of the system with the nucleated domain is greater than the initial reference energy - see~\eqref{nucl_condition}), it follows that nucleation is not energetically favorable at $E_k$ (Fig.~\ref{figure4}).

    \item Next, we consider $E_{va}=E_{va3}<E_k$. The minimum energy point now is N, intersection of (b3'') and $l=d$ (Fig.~\ref{figure3}b), and from \eqref{N_coord} its coordinates are given by:
    \begin{equation} \label{N_Eva3}
        (r^*_{E_{va3}},l^*_{E_{va3}}) = \left(\frac{b}{aE_{va3}},d\right)
    \end{equation}
    As $E_{va3}$ decreases below $E_k$, $r^*_{E_{va3}}$ increases above $r^*_{E_k}$ (blue dashed line in Fig.~\ref{figure4}). From $r^*_{E_{va3}}$ in \eqref{N_Eva3}, we get:
    \begin{equation} \label{E_va3}
        E_{va3} = \frac{b}{ar^*_{E_{va3}}}
    \end{equation}
    Substituting \eqref{N_Eva3} and \eqref{E_va3} into \eqref{U_eq}, we get
    \begin{equation} \label{U_Eva3}
        U^*_{\Delta,E_{va3}} = cr^{*2}_{E_{va3}}
    \end{equation}
    The energy $U^*_{\Delta,E_{va3}}$ increases with increasing $r^*_{E_{va3}}$ (red dashed line in Fig.~\ref{figure4}) and is 0 when $r^*_{E_{va3}}=0<r^*_{E_k}$. This condition is not valid for $E_{va3}<E_k$, since the minimum value of $r^*_{E_{va3}}$ is $r^*_{E_k} >0 $. Hence, there does not exist any $E_{va3}<E_k$ for which nucleation is energetically favorable.

    \item Let us next consider $E_{va}=E_{va1}>E_k$. The minimum energy point now is $M$ (Fig.~\ref{figure3}a), whose coordinates from \eqref{M_coord} are
    \begin{equation} \label{M_Eva1}
        (r^*_{E_{va1}},l^*_{E_{va1}}) = \left( \frac{aE_{va1}d}{2c},d\right)
    \end{equation}
    $r^*_{E_{va1}}$ in \eqref{M_Eva1} leads to
    \begin{equation} \label{Eva1}
        E_{va1} = \frac{2cr^*_{E_{va1}}}{ad}
    \end{equation}
    As $E_{va1}$ increases from $E_k$, $r^*_{E_{va1}}$ increases from $r^*_{E_k}$ (blue solid line in Fig.~\ref{figure4}). Substituting \eqref{M_Eva1} and \eqref{Eva1} in \eqref{U_eq}, we get:
    \begin{equation} \label{U_Eva1}
        U^*{E_{va1}} = bd-cr^{*2}_{E_{va1}}
    \end{equation}
    The energy $U^*_{E_{va1}}$ decreases with increasing $r^*_{E_{va1}}$ (red solid line in Fig~\ref{figure4}) and becomes 0 at $r^*_{E_{va1,c}}$ given by:
    \begin{equation} \label{r_Eva1c}
        r^*_{E_{va1,c}} = \sqrt{\frac{bd}{c}} > r^*_{E_k}
    \end{equation}
    This is a valid point for $E_{va1}>E_k$ and the field $E_{va1,c}$ corresponding to this point can be obtained by substituting \eqref{r_Eva1c} in \eqref{Eva1}, leading to
    \begin{equation} \label{E_vac_eq}
        E_{va,c} = E_{va1,c} = 2\sqrt{\frac{cb}{a^2}}\frac{1}{\sqrt{d}}
    \end{equation}
    This is the critical field $E_{va,c}$ at which nucleation becomes favorable (Fig.~\ref{figure4}). Since, $E_{va,c}>E_k$, favorable nucleation forms a stringent condition over domain growth. Substituting \eqref{E_vac_eq} in \eqref{E_c_E_vac}, we get 
    \begin{equation} \label{Ec_final}
        E_c = 2\sqrt{\frac{cb}{a^2}}\frac{1}{\sqrt{d}} + E_h + \frac{2\sigma_e}{P_s}
    \end{equation}
\end{enumerate}

This result shows that coercive field in FE \ce{HfO2} varies with thickness as $E_c\propto \frac{1}{\sqrt{d}}$. Our analysis shows that this is primarily due to the anisotropic structure of the orthorhombic phase and the resultant half-prolate elliptical cylinder nuclei. This dependence is notably different from the $E_c\propto d^{-2/3}$ as predicted by the JKD theory, which assumes half-prolate spheroid domains.

\section{Experimental Validation of Coercive Field Behavior in \ce{HfO2}}
\label{sec:exp_comp}

\begin{figure}[!h]
\centerline{\includegraphics[width=\columnwidth]{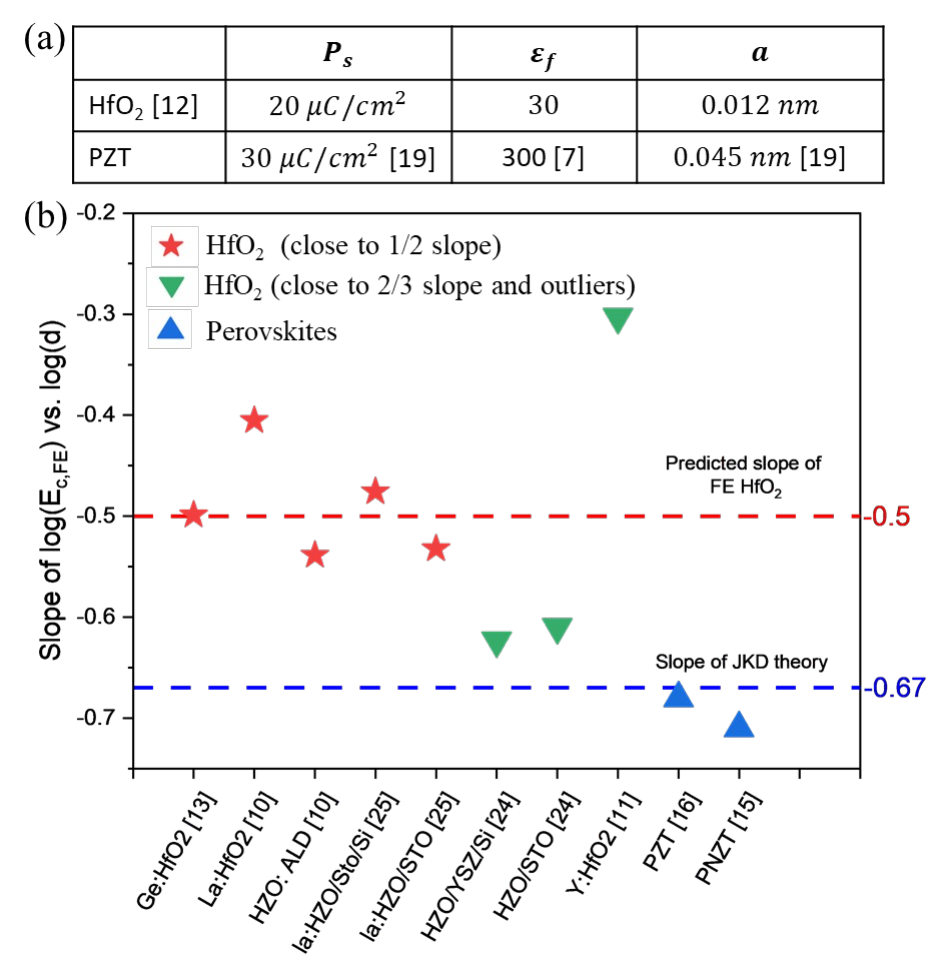}}
\caption{a) Parameter values used to correct experimentally measured coercive field values for depolarization effects in \ce{HfO2} and perovskites. b) Slopes of $\log(E_{c,FE})$ versus. $\log(d)$ of corrected experimental coercive field, showing close to -0.5 slope for FE \ce{HfO2} consistent with the proposed behavior.}
\label{figure5}
\end{figure}

To validate the derived expression for thickness dependence of coercive field in \ce{HfO2}, we compare it with the experimental trends reported in literature across various \ce{HfO2} samples \cite{park_perspective_2023, materano_polarization_2020, mimura_thickness-dependent_2018, toriumi_material_2019, park_study_2015, migita_polarization_2018, fina_epitaxial_2021, song_epitaxial_2020, lyu_growth_2019}. The comparison spans the FE samples with different dopants and processing conditions but is restricted to FE thickness regime of $<100$ nm, where experiments show increase of $E_c$.

In our analysis until now, we have neglected the effect of imperfect screening of polarization charges at the metal-ferroelectric interfaces due to finite screening length of metals \cite{dawber_depolarization_2003, chandra_scaling_2004, park_study_2015}. However, in experimental systems, this interface effect leads to non-negligible depolarization field. In particular, this effect reduces the measured coercive field $(E_{c,meas})$ compared to the intrinsic coercive field $(E_{c,FE})$ of the FE \cite{chandra_scaling_2004}. To correct for this reduction observed in experiments, we include depolarization correction proposed by Dawber et al. \cite{dawber_depolarization_2003} and calculate $(E_{c,FE})$ from $E_{c,meas}$ as follows:

\begin{equation}
    E_{c,FE} = \frac{V_{c,meas}+2P_sa}{d+2a\varepsilon_f}
\end{equation}
here, $V_{c,meas}$ is the experimentally measured coercive voltage (which is $E_{c,meas}\times d$), $d$ is the FE thickness, $P_s$ is the saturation polarization and $\varepsilon_f$ is the permittivity of the FE layer. $a$ is the normalized screening length of the metal which is $\lambda/\varepsilon_e$, $\lambda$ is the screening length and $\varepsilon_e$ is the metal permittivity.

Using the parameters shown in Fig.~\ref{figure5}a, we calculate $E_{c,FE}$ for the \ce{HfO2} samples. For comparative analysis, we also include the experimental data for perovskites \cite{bjormander_thickness_1995,pertsev_coercive_2003}. We utilize the $E_{c,FE}$ values across $d$ and calculate the slope of $\log(E_{c,FE})$ versus $\log(d)$ using a robust linear regression fit to reduce the effect of outliers. The resulting slope values for different \ce{HfO2} experimental samples (Fig.~\ref{figure5}) show good agreement with our prediction of 1/2 slope. In contrast, perovskites exhibit a slope of 2/3. It is noteworthy that some epitaxial samples of \ce{HfO2} also show a slope close to 2/3, which can be due to the prevalence of the ferroelectric rhombohedral phase (\ce{{R3mn}} - exhibiting the absence of APSL and confined domain growth) rather than orthorhombic phase, as discussed earlier.

\section{Conclusion}
We theoretically demonstrate that the reduced thickness dependence of coercive field in ferroelectric \ce{HfO2} (compared to perovskites) stems from the anisotropic crystal structure of the polar orthorhombic phase of \ce{HfO2}. Specifically the presence of alternating polar-spacer layers (APSL) in one lateral direction leads to high energy barrier for domain propagation constraining the domain growth. As a result, nucleated domains favorably expand along the continuous polar layer (CPL) direction, leading to half-prolate elliptical cylindrical domain geometry rather than the convention half-prolate spheroid domains. By analyzing the energetics of formation and growth of these half-prolate elliptical cylindrical domains, we derive a modified scaling relation for coercive field as $E_c\propto d^{-1/2}$. These revised exponent of 1/2 agrees well with the experimental results and captures the effect of confined domain growth in \ce{HfO2}.

\bibliographystyle{IEEEtran}
\bibliography{MFM_Ec_TFE.bib}

\end{document}